\documentstyle[twoside,psfig,amsfonts]{article}

\catcode`\@=11
\long\def\@makefntext#1{
\protect\noindent \hbox to 3.2pt {\hskip-.9pt  
$^{{\eightrm\@thefnmark}}$\hfil}#1\hfill}               

\def\@makefnmark{\hbox to 0pt{$^{\@thefnmark}$\hss}}    
        
\def\ps@myheadings{\let\@mkboth\@gobbletwo
\def\@oddhead{\hbox{}
\rightmark\hfil\eightrm\thepage}   
\def\@oddfoot{}\def\@evenhead{\eightrm\thepage\hfil
\leftmark\hbox{}}\def\@evenfoot{}
\def\sectionmark##1{}\def\subsectionmark##1{}}

\oddsidemargin=\evensidemargin
\newcounter{sectionc}\newcounter{subsectionc}\newcounter{subsubsectionc}
\renewcommand{\section}[1] {\vspace{12pt}\addtocounter{sectionc}{1} 
\setcounter{subsectionc}{0}\setcounter{subsubsectionc}{0}\noindent 
        {\tenbf\thesectionc. #1}\par\vspace{5pt}}
\renewcommand{\subsection}[1] {\vspace{12pt}\addtocounter{subsectionc}{1} 
        \setcounter{subsubsectionc}{0}\noindent 
        {\bf\thesectionc.\thesubsectionc. {\kern1pt \bfit #1}}\par\vspace{5pt}}
\renewcommand{\subsubsection}[1] {\vspace{12pt}\addtocounter{subsubsectionc}{1}
        \noindent{\tenrm\thesectionc.\thesubsectionc.\thesubsubsectionc.
        {\kern1pt \tenit #1}}\par\vspace{5pt}}

\newcounter{appendixc}
\newcounter{subappendixc}[appendixc]
\newcounter{subsubappendixc}[subappendixc]
\renewcommand{\thesubappendixc}{\Alph{appendixc}.\arabic{subappendixc}}
\renewcommand{\thesubsubappendixc}
        {\Alph{appendixc}.\arabic{subappendixc}.\arabic{subsubappendixc}}

\renewcommand{\appendix}[1] {\vspace{12pt}
        \refstepcounter{appendixc}
        \setcounter{figure}{0}
        \setcounter{table}{0}
        \setcounter{lemma}{0}
        \setcounter{theorem}{0}
        \setcounter{corollary}{0}
        \setcounter{definition}{0}
        \setcounter{equation}{0}
        \renewcommand{\thefigure}{\Alph{appendixc}.\arabic{figure}}
        \renewcommand{\thetable}{\Alph{appendixc}.\arabic{table}}
        \renewcommand{\theappendixc}{\Alph{appendixc}}
        \renewcommand{\thelemma}{\Alph{appendixc}.\arabic{lemma}}
        \renewcommand{\thetheorem}{\Alph{appendixc}.\arabic{theorem}}
        \renewcommand{\thedefinition}{\Alph{appendixc}.\arabic{definition}}
        \renewcommand{\thecorollary}{\Alph{appendixc}.\arabic{corollary}}
        \renewcommand{\theequation}{\Alph{appendixc}.\arabic{equation}}
        \noindent{\tenbf Appendix \theappendixc #1}\par\vspace{5pt}}
\newcommand{\subappendix}[1] {\vspace{12pt}
        \refstepcounter{subappendixc}
        \noindent{\bf Appendix \thesubappendixc. {\kern1pt \bfit #1}}
        \par\vspace{5pt}}
\newcommand{\subsubappendix}[1] {\vspace{12pt}
        \refstepcounter{subsubappendixc}
        \noindent{\rm Appendix \thesubsubappendixc. {\kern1pt \tenit #1}}
        \par\vspace{5pt}}

\newcommand{\textlineskip}{\baselineskip=13pt}
\newcommand{\smalllineskip}{\baselineskip=10pt}

\def\eightcirc{
\begin{picture}(0,0)
\put(4.4,1.8){\circle{6.5}}
\end{picture}}
\def\eightcopyright{\eightcirc\kern2.7pt\hbox{\eightrm c}}

\def\abstracts#1#2#3{{
        \centering{\begin{minipage}{4.5in}\footnotesize\baselineskip=10pt
        \parindent=0pt #1\par 
        \parindent=15pt #2\par
        \parindent=15pt #3
        \end{minipage}}\par}} 

\newcommand{\bibit}{\nineit}
\newcommand{\bibbf}{\ninebf}
\renewenvironment{thebibliography}[1]
        {\frenchspacing
         \ninerm\baselineskip=11pt
         \begin{list}{\arabic{enumi}.}
        {\usecounter{enumi}\setlength{\parsep}{0pt}     
         \setlength{\leftmargin 12.7pt}{\rightmargin 0pt} 
         \setlength{\itemsep}{0pt} \settowidth
        {\labelwidth}{#1.}\sloppy}}{\end{list}}

\newcounter{itemlistc}
\newcounter{romanlistc}
\newcounter{alphlistc}
\newcounter{arabiclistc}

\newcommand{\fcaption}[1]{
        \refstepcounter{figure}
        \setbox\@tempboxa = \hbox{\footnotesize Fig.~\thefigure. #1}
        \ifdim \wd\@tempboxa > 5in
           {\begin{center}
        \parbox{5in}{\footnotesize\smalllineskip Fig.~\thefigure. #1}
            \end{center}}
        \else
             {\begin{center}
             {\footnotesize Fig.~\thefigure. #1}
              \end{center}}
        \fi}

\newcommand{\tcaption}[1]{
        \refstepcounter{table}
        \setbox\@tempboxa = \hbox{\footnotesize Table~\thetable. #1}
        \ifdim \wd\@tempboxa > 5in
           {\begin{center}
        \parbox{5in}{\footnotesize\smalllineskip Table~\thetable. #1}
            \end{center}}
        \else
             {\begin{center}
             {\footnotesize Table~\thetable. #1}
              \end{center}}
        \fi}

\def\@citex[#1]#2{\if@filesw\immediate\write\@auxout
        {\string\citation{#2}}\fi
\def\@citea{}\@cite{\@for\@citeb:=#2\do
        {\@citea\def\@citea{,}\@ifundefined
        {b@\@citeb}{{\bf ?}\@warning
        {Citation `\@citeb' on page \thepage \space undefined}}
        {\csname b@\@citeb\endcsname}}}{#1}}

\newif\if@cghi
\def\cite{\@cghitrue\@ifnextchar [{\@tempswatrue
        \@citex}{\@tempswafalse\@citex[]}}
\def\citelow{\@cghifalse\@ifnextchar [{\@tempswatrue
        \@citex}{\@tempswafalse\@citex[]}}
\def\@cite#1#2{{$\null^{#1}$\if@tempswa\typeout
        {IJCGA warning: optional citation argument 
        ignored: `#2'} \fi}}

\def\pmb#1{\setbox0=\hbox{#1}
        \kern-.025em\copy0\kern-\wd0
        \kern.05em\copy0\kern-\wd0
        \kern-.025em\raise.0433em\box0}

\def\fnt#1#2{\footnotetext{\kern-.3em
        {$^{\mbox{\scriptsize #1}}$}{#2}}}

\def\fpage#1{\begingroup
\voffset=.3in
\thispagestyle{empty}\begin{table}[b]\centerline{\footnotesize #1}
        \end{table}\endgroup}

\def\runninghead#1#2{\pagestyle{myheadings}
\markboth{{\protect\footnotesize\it{\quad #1}}\hfill}
{\hfill{\protect\footnotesize\it{#2\quad}}}}
\font\tenrm=cmr10
\font\tenit=cmti10 
\font\tenbf=cmbx10
\font\bfit=cmbxti10 at 10pt
\font\ninerm=cmr9
\font\nineit=cmti9
\font\ninebf=cmbx9
\font\eightrm=cmr8

\def\qed{\hbox{${\vcenter{\vbox{                        
   \hrule height 0.4pt\hbox{\vrule width 0.4pt height 6pt
   \kern5pt\vrule width 0.4pt}\hrule height 0.4pt}}}$}}

\newcommand{\cruza}{\!\not\!}
\def\slash#1{\raise.18ex\hbox{/}\kern-.50em #1}
\def\bbox#1{{ \mbox{\boldmath $#1$}} }

\begin{document}
\setlength{\textheight}{7.7truein}  

\runninghead{J.L. Alonso et al.}{Mass 
protection via translational invariance}

\normalsize\textlineskip
\thispagestyle{empty}
\setcounter{page}{1}



\fpage{1}
\centerline{\bf MASS PROTECTION VIA TRANSLATIONAL INVARIANCE}
\baselineskip=13pt
\vspace*{0.37truein}
\centerline{\footnotesize JOS\'E LUIS 
ALONSO\footnote{E-mail: buj@gteorico.unizar.es} \, and JOS\'E LUIS 
CORT\'ES\footnote{E-mail: cortes@leo.unizar.es}}
\baselineskip=12pt
\centerline{\footnotesize\it Departamento de F\'{\i}sica Te\'orica, 
Universidad de Zaragoza, 50009 Zaragoza, Spain}
\vspace*{10pt}

\centerline{\footnotesize PHILIPPE 
BOUCAUD\footnote{E-mail: phi@qcd.th.u-psud.fr}}
\baselineskip=12pt
\centerline{\footnotesize\it Laboratoire de Physique Th\'eorique 
et Hautes \'Energies, Universit\'e de Paris XI,}
\baselineskip=10pt
\centerline{\footnotesize\it 91405 Orsay, France}
\vspace*{10pt}

\centerline{\footnotesize JOS\'E MANUEL 
CARMONA\footnote{E-mail: carmona@difi.unipi.it}}
\baselineskip=12pt
\centerline{\footnotesize\it Dipartimento di Fisica, Universit\'a di Pisa,
Via Buonarroti, 2, 56127 Pisa, Italy}
\vspace*{10pt}

\centerline{\footnotesize JANOS 
POLONYI\footnote{E-mail: polonyi@fresnel.u-strasbg.fr}}
\baselineskip=12pt
\centerline{\footnotesize\it Laboratoire de Physique Th\'eorique, 
Universit\'e Louis Pasteur,}
\baselineskip=10pt
\centerline{\footnotesize\it 67084 Strasbourg CEDEX, France}
\centerline{\footnotesize\it and}
\centerline{\footnotesize\it Department of Atomic Physics, 
L. E\"otv\"os University, Budapest, Hungary}
\vspace*{10pt}

\centerline{\footnotesize ARJAN VAN DER 
SIJS\footnote{E-mail: arjan@scsc.ethz.ch}}
\baselineskip=12pt
\centerline{\footnotesize\it Swiss Center for Scientific Computing, 
ETH Z\"urich, ETH-Zentrum,} 
\baselineskip=10pt
\centerline{\footnotesize\it CH-8092 Z\"urich, Switzerland}
\vspace*{10pt}

\vspace*{0.1truein}


\vspace*{0.21truein}
\abstracts{We propose a way of protecting a Dirac fermion interacting
with a scalar field from acquiring a mass from the vacuum. It is obtained
through an implementation of  
translational symmetry when the theory is formulated with a momentum cutoff,
which forbids the usual Yukawa term.
We consider that this mechanism can help to understand the
smallness of neutrino masses  without a tuning of the Yukawa coupling.
The prohibition of the Yukawa term for the neutrino 
forbids at the same time a gauge coupling  between the right-handed 
electron and neutrino.
We prove that this mechanism can be implemented on the lattice.}{}{}



\medskip
\vfill


\vspace*{1pt}\textlineskip      
\section{Introduction}  
\vspace*{-0.5pt}
\noindent
In the minimal standard model  a right-handed chirality for the neutrino is
absent, left-handed leptons are in weak SU(2) doublets
and right-handed leptons in weak SU(2) singlets.
 If a $\nu_R$ does exist, then 
 the neutrino could get a Dirac mass 
$m_{\nu_e}(\bar \nu_{e_L} \nu_{e_R} + \mathrm{h.c.})$ and
a fundamental problem
is to understand why $m_{\nu_e}/m_e$ is such a small number ($<10^{-5}$).
The standard answer to this problem is the see-saw mechanism, which is 
based on the introduction of a Majorana mass term for the right-handed
neutrino with a high enough new mass scale.$^1$

We explore in this letter another possible cause for the smallness of
neutrino masses, which is 
offered by a new implementation of translational symmetry
in the context of the Standard Model (SM) as an effective theory.

To be more specific, the SM is a theory where
first, the symmetries are postulated, and second, the representations in 
which the elementary particles appear are chosen. This method has been 
explicitly implemented for the gauge symmetries and the little group of the 
Lorentz group, classifying particles into different representations of these
groups. The different possible representations 
 of the group of translations have not been relevant up to
now (to our knowledge) in the formulation of the SM. It is in this context
that we make our proposal. We will show that the choice 
 of a different representation for the left and right-handed fermion
fields offers the possibility to forbid the usual Yukawa mass term.
We remark that the choice of the
representations will be justified (at least for the moment) simply by the 
phenomenological results of the theory. 

\section{Mass protection mechanism}
\noindent
To illustrate how this mechanism works, 
we will consider a chiral model with  a  left  and a right fermion coupled 
to a complex scalar field.

Setting
\begin{equation}
\phi(x)=\phi_{1}(x)+i\phi_{2}(x),
\end{equation}
 we will consider that  the scalar field $\phi$ gets 
a VEV $\langle\phi_{1}(x)\rangle=v$, $\langle\phi_{2}(x)\rangle=0$. 
Both the scalar action and the vacuum are translationally invariant when
 $\phi$ transforms
in the usual way under a translation by a vector~$r$, 
\begin{equation}
\phi'(x)=\phi(x+r).
\label{usual}
\end{equation}

In order to identify the new representation for translations 
it is necessary to introduce a momentum cutoff 
($-\Lambda \leq p_\mu \leq \Lambda$), which is a natural 
way of incorporating the limitations on the domain of 
validity of the model, which we consider as a low energy
effective theory. As a consequence of the introduction 
of the cutoff, the group O(4) (Euclidean version of the
homogeneous Lorentz group) is replaced by the discrete
subgroup generated by rotations of angle $\pi/2$ in
each plane, and the group of translations is replaced by the
discrete subgroup generated by translations of $\pi/\Lambda$
in each direction.

Now when one considers the possible linear representations of 
the discretized translations ($r^\mu = n^\mu \pi/\Lambda$ ,
$n^\mu \in \Bbb{Z}$) for the fermion field one has \textit{only} two different 
choices which are \textit{compatible with the usual representations of
rotations,}
\begin{equation}
\psi'(p) = e^{ir.p}\psi(p),
\label{rep1}
\end{equation}
which is the momentum space version of the usual transformation
law~(\ref{usual}) and a new representation,
\begin{equation}
\psi'(p) = e^{ir.{\tilde p}}\psi(p)
\label{rep2}
\end{equation}
where ${\tilde p}_\mu = p_\mu -\Lambda \epsilon(p_\mu)$, and 
$\epsilon(p_\mu)=\mathrm{sign}(p_\mu)$.

In coordinate space, the representation~(\ref{rep2}) reads simply
\begin{equation}
\psi'(x)=e^{i\Lambda\sum_\mu r_\mu}\psi(x+r).
\end{equation}
Note that for an elementary
translation, the difference between Eqs.~(\ref{rep1}) and~(\ref{rep2})
is just a minus sign, though this sign will lead to very relevant consequences 
as we will see.

In the context of effective theories,$^2$ the new representation
of translations could offer the possibility to describe remnants at 
low energy of high energy phenomena which cannot be described in terms
of higher dimensional operators, like inhomogeneous vacua for example.
At the end of this section we comment a little more on a possible 
interpretation of this new representation of translations.

In order to get any consequences from the use of the new 
representation in an action build out of bilinears in the fermion 
field one has to consider a different representation for each 
chirality. In this case, the usual Yukawa term in momentum space,
\begin{equation}
y \bar\psi_L([p+k])\phi(k)\psi_R(p),
\label{usualyukawa}
\end{equation}
is forbidden by translational invariance. In Eq.~(\ref{usualyukawa}),
$[p+k]$ is the momentum compatible with the cutoff obtained 
by adding or substracting if necessary $2\Lambda$ to the components 
of $p+k$. The interaction 
term compatible with the new implementation of translations is
\begin{equation} 
y \bar\psi_L(\widetilde{[p+k]})\phi(k)\psi_R(p),
\label{newyukawa}
\end{equation}
where the tilde symbol was already introduced in Eq.~(\ref{rep2}).

Now one can compare the results for the fermion propagator in the
two cases: either with the same transformation law under translations for both
chiralities (which implies the usual Yukawa interaction) or with different
representations for each chirality (in this case, the appropriate Yukawa 
term is Eq.~(\ref{newyukawa})).

In the approximation where the fluctuations of the scalar field are
neglected, one has, in the usual case, the propagator of a free fermion
with mass $m=yv$. In the case of the new translation representation,
one finds the propagator of a massless fermion up to corrections 
proportional to inverse powers of the cutoff $\Lambda$. So we can say
that translational invariance can protect a Dirac fermion
from acquiring a mass from the vacuum.

However, as the term~(\ref{newyukawa}) couples momentum modes that 
differ in $\Lambda$, a nonperturbative implementation of this 
mechanism could be problematic owing to the well-known fermion
doubling phenomenon.
We will later see that this is not the case.

Coming back to possible physical interpretations of the new representation
of translations, one can note that the very existence of 
different representations
for the left- and right-handed parts of a field comes from the presence
of a cutoff. One could interpret this by saying that
the chirality transforming in the new representation will have a different
coupling to the physics beyond the cutoff from that of the chirality
that transforms in the usual way under translations.

\section{Application to the SM}
\noindent
This mechanism could be applied to understand the mass difference
between the electron and
the neutrino in the SM (or, in general, between a charged lepton and
its corresponding neutrino)
by assuming that the right-handed neutrino
transforms differently from the left-handed neutrino under
translations.  We remark
that we do not introduce a new symmetry, but only use a new representation
of an existing symmetry, translational invariance.

To apply this mechanism in the case of the electron and the neutrino,
we first have to remember that the
left-handed electron and the left-handed neutrino form an SU(2)
doublet, so they are coupled by the gauge field in terms such as
$\bar e_L\gamma^{\mu}W_{\mu}\nu_L$. In order to leave these terms 
translationally invariant with the gauge field transforming in the trivial
representation of translations, we need the same representation of 
translations for both the left-handed electron field and the left-handed
neutrino field.
Besides, we want to give mass to the electron with a usual Yukawa term
through the Higgs mechanism,
so we have to choose the same representation for the two chiral components
of the electron field.
For the neutrino field, we  take 
differents representations for the two chiral components
to forbid the usual Yukawa term. Then the right-handed electron field and the
right-handed neutrino field are in different representations and they
cannot be in the same weak isospin multiplet.
This situation is in fact assumed in the SM.

  Majorana terms such as
\begin{equation}
\psi^{\mathrm{T} }_{L}(x) C \psi^{}_{L}(x)\phi(x)\phi(x) \, ,
\quad
\psi^{\mathrm{T} }_{R}(x) C \psi_{R}(x)
\end{equation}
(where $C$ is the charge-conjugation operator) are compatible with
translational invariance independently of the choice of representations
and should be included in a more detailed analysis of neutrino masses.

\section{Lattice implementation}
\noindent
As we have previously said, the term~(\ref{newyukawa}) couples momentum modes
that differ in $\Lambda$ so that one could be afraid of possible problems
in a nonperturbative implementation of our proposal. In fact, a lattice
field theory is a good starting point to derive properties of the theory
in a rigorous way, but it suffers from the illness of the fermion doubling 
phenomenon. As it is well known, this illness is more serious when the two
chiralities of a fermion are not treated on the same footing. Therefore,
it is pertinent to control that the mechanism we have proposed 
is compatible with a lattice formulation, in the sense that the lattice
doublers still decouple, while the unconventional choice of representation
of the translational symmetry for the lattice fermion field protects the
generation of a fermion mass to all orders in perturbation theory.

The lattice action for the
scalar field is (we will take the lattice spacing $a=1$)~:
\begin{eqnarray}
S_{\mathrm{B}}&=&-\kappa\sum_{x,\mu}
(\phi^*_x\phi_{x+\hat\mu}+\phi^*_{x+\hat\mu}\phi_x) \nonumber \\
&&+\sum_x\{\phi^*_x\phi_x+
\lambda(\phi^*_x\phi_x-1)^2\} .
\end{eqnarray}

On the lattice, we take for the  
representation of translations for the fermion field~: 
\begin{equation}
\psi'_{Lx}=e^{i\alpha_L}\psi_{L x+\hat\mu}\, ,\quad
\psi'_{Rx}=e^{i\alpha_R}\psi_{R x+\hat\mu},
\label{newreplat}
\end{equation}
under a translation of one lattice spacing 
in the $\hat\mu$ direction. As in the previous discussion, 
in order to have compatibility with the usual
representations of rotations, only 0 and $\pi$ will be the allowed values
for $\alpha_L$, $\alpha_R$ in Eq.~(\ref{newreplat}). 
Then the lattice na\"\i ve kinetic term~: 
\begin{equation}
S_{\mathrm{F}}=\sum_{x,\mu}\frac{1}{2}(\bar\psi_x\gamma_\mu\psi_{x+\hat\mu}-
\bar\psi_{x+\hat\mu}\gamma_\mu\psi_x),
\label{kinetic}
\end{equation}
is rotational and translational invariant for any of the two allowed values 
for $\alpha_L$ and $\alpha_R$ and the Yukawa term is  forbidden
when $\alpha_L \ne \alpha_R$.
But with the new behaviour under translations, we 
have to admit coupling constants which depend on the space coordinates.
With the choice
\begin{equation}
\alpha_R=\pi,\quad \alpha_L=0 ,
\label{condition}
\end{equation}
a term of the following type~: 
\begin{equation}
y\sum_x (-1)^{\sum_\nu x_\nu}
(\bar\psi_{Lx}\phi_x\psi_{Rx}+
\bar\psi_{Rx}\phi_x^*\psi_{Lx})
\end{equation}
is allowed by the symmetry. 

A difference between the continuum and the lattice discussion is that
 the na\"\i ve lattice kinetic term
(\ref{kinetic}) suffers from the doubling phenomenon.
To avoid  this last problem,  we will follow
the methods of Refs. 3 and 4 and write
the  boson-fermion interaction as~:
\begin{equation}
S_{\mathrm{FB}}=y\sum_x (-1)^{\sum_\nu x_\nu}
(\bar\psi^{(1)}_{Lx}\phi_x\psi^{(1)}_{Rx}+
\bar\psi^{(1)}_{Rx}\phi_x^*\psi^{(1)}_{Lx}),
\label{newterm}
\end{equation}
where 
\begin{equation}
\psi^{(1)}(p)=F(p)\psi(p).
\end{equation}
$F(p)$ is a form factor required to be 1 for $p=0$ and
to vanish when $p$ equals any of the doubler momenta. 
With this method, we have a theory with 16 fermions, 15 of which do not
interact with physical particles and decouple from the real world. 

We will show now that the translationally invariant action 
including the new term $S_{FB}$ (\ref{newterm}), 
\begin{equation}
S=S_{\mathrm{B}}+S_{\mathrm{F}}+S_{\mathrm{FB}},
\label{action}
\end{equation}
gives a zero mass for the fermion.

Let us first note that the presence in the action of the 
term $S_{\mathrm{FB}}$ with such an unusual coupling does not modify the vacuum
$\langle\phi_{1x}\rangle=v$.  This is a consequence of both analytical and
numerical studies of the antiferromagnetic (AFM) phase of the chiral Yukawa
model.$^5$ Under the change of variables 
$\phi'_x=\varepsilon_x\phi_x$, where $\varepsilon_x=(-1)^{\sum_\nu x_\nu}$, 
the action is invariant if the couplings are mapped according to
\begin{equation}
\left(\kappa,y\varepsilon_x\right)\longmapsto
\left(-\kappa,y\right).
\end{equation}
With these couplings, a stable AFM phase exists where the scalar gets a 
staggered mean value $\langle\phi'_{1x}\rangle=\varepsilon_x
v_{\mathrm{st}}$. We can then  conclude that the original vacuum
$\langle\phi_{1x}\rangle=v$ is  also a  stable  vacuum for the action
 (\ref{action}).

In order to do perturbation theory, let us rewrite $S_{\mathrm{FB}}\,$ 
in the form
\begin{equation}
S_{\mathrm{FB}}=y\sum_x \varepsilon_x
(\bar\psi^{(1)}_x \{\phi_x P_R+\phi^*_x P_L\}\psi^{(1)}_x),
\end{equation}
where $P_L=\frac{1}{2}(1-\gamma_5)$, $P_R=\frac{1}{2}(1+\gamma_5)$. 

 We write
\begin{equation}
\phi_{1x}=v+\eta_{1x} \, , \quad \phi_{2x}=\eta_{2x},
\label{bosons}
\end{equation}
where $\eta_{1,2}$ represent the small perturbations. In momentum space,
the fermionic matrix at tree-level order is
\begin{equation}
i\delta(p-p')\cruza s(p)+
 y v F(p)F_\pi(p)\delta(p-p'+\bbox{\pi}),
\end{equation}
where $\slash{s}(p)=\sum_\mu \gamma_\mu \sin p_\mu$, 
$F_\pi(p)\equiv F(p+\bbox{\pi})$, and $\bbox{\pi}\equiv(\pi,\pi,\pi,\pi)$. 
We have $F_\pi(0)=0.$
This matrix is
not diagonal in momentum space, as it connects every momentum $p$
with $p+\bbox{\pi}$ in a box of the form
\begin{equation}
G^{-1}(p)=\left(\begin{array}{cc}
i\cruza s(p) &  y v F(p)F_\pi(p) \\
 y v F_\pi(p)F(p) & i\cruza s(p+\bbox{\pi})
\end{array}\right).
\label{invprop}
\end{equation}
It can be diagonalized with the non-unitary transformation
\begin{equation}
G^{-1}_{\mathrm{D}}(p)=TG^{-1}(p)T, \quad 
T=\frac{1}{\sqrt{2}}\left(\begin{array}{cc}
1 & i \\ i & 1\end{array}\right),
\label{diag1}
\end{equation}
to give
\begin{equation}
G_{\mathrm{D}}=\frac{1}{i\cruza s(p)+im(p)}\, ,
\label{diagprop}
\end{equation}
where $m(p)= y v F(p)F_\pi(p)=m(p+\bbox{\pi})$. This
propagator has 16 poles at momenta $(0,0,0,0)$, $(\pi,\pi,\pi,\pi)$,
$(\pi,0,0,0)$, $(\pi,\pi,0,0)$, etc., which implies zero mass at tree level for
the physical fermion and all the doublers.
A straightforward mean field 
calculation$^4$ leads to the same conclusion
but with a contribution to $v$ coming from the fermionic condensate besides
the contribution of the boson VEV.

Now we want to know if this masslessness is maintained when 
loop corrections are added. From the interaction term with the bosons
$\eta_1$ and $\eta_2$ defined in Eq.~(\ref{bosons}),
\begin{eqnarray}
 y \int^\pi_{-\pi}\frac{d^4p}{(2\pi)^4}
\frac{d^4p'}{(2\pi)^4}
\bar\psi(p)F(p) \{\eta_1(p-p'+\bbox{\pi})\nonumber \\
+\eta_2(p-p'+\bbox{\pi})i\gamma_5\} F(p')\psi(p'),
\end{eqnarray}
we get the Feynman rules of interest~: a 
fermion-boson-fermion vertex which inserts a momentum  $\bbox{\pi}$, and
contributions coming from the diagonal and nondiagonal (connecting
momenta $k$ and $k+\bbox{\pi}$) elements of the propagator matrix. 
One can then calculate the
1-loop fermion self-energy diagram, which is then added to the inverse
propagator to give
\begin{equation}
G^{-1}_{\mathrm{1-loop}}(p)
=\left(\begin{array}{cc}
i\cruza s(p)+\Sigma^A(p) & m(p)+\Sigma^B(p) \\
m(p)+\Sigma^B_\pi(p) & -i\cruza s(p)+\Sigma^A_\pi(p)
\end{array}\right),
\label{1loopprop}
\end{equation}
where $\Sigma^{A,B}_\pi(p) \equiv \Sigma^{A,B}(p+\bbox{\pi})$, and
\begin{eqnarray}
\Sigma^A(p)&=& y^2 F^2(p)\int_{-\pi}^\pi \frac{d^4 k}{(2\pi)^4}
F^2(k) \frac{-i\cruza s(k)}{s^2(k)-m^2(k)} \nonumber \\
&&\times\{{\cal G}_1(p-k+\bbox{\pi})+{\cal G}_2(p-k+\bbox{\pi})\},
\end{eqnarray}
\begin{eqnarray}
&&\Sigma^B(p)= y^2F(p)F_\pi(p)\int_{-\pi}^\pi \frac{d^4 k}{(2\pi)^4}
F(k)F_\pi(k) \nonumber \\
&&\times\frac{-m(k)}{s^2(k)-m^2(k)} 
\{{\cal G}_1(p-k+\bbox{\pi})-{\cal G}_2(p-k+\bbox{\pi})\},
\end{eqnarray}
${\cal G}_j$ being the  $\eta_j$ boson  propagator $(j=1,2)$~:
\begin{equation}
{\cal G}_j(q)=\frac{(2\kappa)^{-1}}{2\sum_\rho (1-\cos q_\rho)+m_j^2}\, ,
\label{bosonprop}
\end{equation}
with $m_1=(4\lambda/\kappa)v^2$  and  $m_2=0$.
It is easy to see that $\Sigma^B_\pi(p)=\Sigma^B(p)$.

The new propagator (\ref{1loopprop}) cannot be diagonalized with
the transformation (\ref{diag1}). But in fact there is a general 
diagonalization that can be used for both the tree-level and the one-loop
inverse propagators, (\ref{invprop}) and (\ref{1loopprop}) respectively.
The transformation is
\begin{equation}
U G^{-1}V, \, \,
U=\left(\begin{array}{cc}
P_L & iP_R \\
i P_R & P_L\end{array}\right),\,\, 
V=\left(\begin{array}{cc}
P_R & iP_L \\
i P_L & P_R\end{array}\right).
\end{equation}

Actually, it is easy to convince oneself  that the matrices $U$
and $V$ will diagonalize the propagator at every loop order~:  
notice that the change of field variables given by these 
two matrices, which in coordinate space is written as
\begin{eqnarray}
\psi_{Rx}=\chi_{Rx} \quad &,& \quad \bar\psi_{Rx}=\bar\chi_{Rx},\\
\psi_{Lx}=i \varepsilon_x\chi_{Lx} \quad &,& \quad
\bar\psi_{Lx}=i \varepsilon_x \bar\chi_{Lx},
\end{eqnarray}
leaves  the kinetic term invariant and changes the term (\ref{newterm}) 
into
\begin{equation}
i y \sum_x (\bar\chi^{(16)}_{Lx}\phi_x\chi^{(1)}_{Rx}+
\bar\chi^{(1)}_{Rx}\phi_x^*\chi^{(16)}_{Lx}),
\label{termdoublers}
\end{equation}
where
\begin{equation}
\chi^{(16)}(p)=F_\pi(p)\chi(p)
\end{equation}
selects the fermion centered around the doubler momentum 
$\bbox{\pi}$.$^4$ In terms of the new variables 
$\chi$ we have a coupling that does not depend on the space coordinates,
so that in momentum space the propagator will be diagonal at every
order in perturbation theory. 

$\Sigma^B$ is proportional to   $F(p)F_\pi(p)$, so it vanishes
at $p=0$ and at all doubler momenta.
$\Sigma^A$ is also zero for the same values because 
it is proportional to  $F(p)$  and for $p=0$ it is the integral of 
an odd function.
In conclusion, the position of the poles is not modified
by the one-loop corrections, therefore both the physical 
fermion and the doublers have zero masses. 

This is  also true  when one investigates two-loop diagrams.
In fact, there is a non-perturbative argument that shows that 
the chirally nondiagonal
 part of the proper vertex
with two external legs (i.e. the inverse of the pro\-pa\-ga\-tor) is zero when
evaluated at any doubler momentum, including $p=0,\bbox{\pi}$.
As discussed in Ref.~3, there is a symmetry of the action 
under appropriate  (``shift'') transformations of the fermion field,
which, when applied to the term (\ref{termdoublers}), 
leads to Ward identities 
 which express that
\begin{equation}
\Gamma^{(2)}_{\mathrm{LR}}(p)=0 
\end{equation}
when evaluated at $p$ equal to one of the doubler momenta. This is
a consequence of the decoupling method together with the form of the
term (\ref{termdoublers}), and 
it is responsible for the cancellation of $\Sigma^B$.

\section{Conclusion}
\noindent
We have shown that the freedom in the choice of the
representations of the translational symmetry can provide  a  mechanism
 which imposes a great restriction on the generation of the  mass for a 
fermion.
In particular, one could understand
the  absence of a Dirac mass contributions for the neutrino.
With this mechanism, the prohibition of the Yukawa term for the neutrino 
forbids at the same time a gauge coupling  between the right-handed 
electron and neutrino.

We are grateful to Fernando Falceto for several useful comments.
This work was supported by CICYT (Spain) projects
AEN-97-1680, AEN-97-1708 and the Actions Int\'egr\'ees
franco-espagnoles HF1996-0022, HF1997-0041. J.M.C.
thanks the Spanish MEC, CAI European 
program, DGA (CONSI+D) and the EU (TMR Network no. FMRX-CT97-0122) 
for financial support.

\end{document}